\documentstyle[12pt]{article}
\begin{document}
%------------------------------------------------------------------------------
%MARCO FOR ABSTRACT BLOCK
\def\abstracts#1{{
	\centering{\begin{minipage}{30pc}\tenrm\baselineskip=12pt\noindent
	\centerline{\tenrm ABSTRACT}\vspace{0.3cm}
	\parindent=0pt #1
	\end{minipage} }\par}}

%------------------------------------------------------------------------------
%NEW MACRO FOR BIBLIOGRAPHY
\newcommand{\bibit}{\it}
\newcommand{\bibbf}{\bf}
\renewenvironment{thebibliography}[1]
	{\begin{list}{\arabic{enumi}.}
	{\usecounter{enumi}\setlength{\parsep}{0pt}
%1.25cm IS STRICTLY FOR PROCSLA.TEX ONLY
\setlength{\leftmargin 1.25cm}{\rightmargin 0pt}
%0.52cm IS FOR NEW DATA FILES
%\setlength{\leftmargin 0.52cm}{\rightmargin 0pt}
	 \setlength{\itemsep}{0pt} \settowidth
	{\labelwidth}{#1.}\sloppy}}{\end{list}}

%----------------------START OF DATA FILE------------------------------

\centerline{\tenbf LIGHT QUARK MASSES AND MIXING ANGLES}

\vspace{0.8cm}
\centerline{\tenrm JOHN F. DONOGHUE}
\baselineskip=13pt
\centerline{\tenit Department of Physics and Astronomy, University
of Massachusetts }
\baselineskip=12pt
\centerline{\tenit Amherst MA 01002 USA}
\vspace{0.9cm}
\abstracts{
I review the present state of our knowledge about the masses and weak
mixing elements of the u, d, s quarks. This is the written version of
lectures given at the 1993 Theoretical Advanced Study Institute (TASI).
}

\vfil
%\vspace{0.8cm}
%\twelverm
\baselineskip=14pt
\section{Introduction}

The Standard Model is clearly one of the triumphs of modern science.
However one of the less pleasant aspects of the theory is that it contains so
many free parameters.  Some of these parameters form the topic of these
lectures,
namely the masses $m_u , m_d$ and $m_s$ and the weak mixing elements
$V_{ud}$ and $V_{us}$.  Within the model, all are products of the Higgs
sector.  They seem to be almost arbitrary numbers, but perhaps they are
clues as to the structure beyond the Standard Model.  Perhaps
someday we will learn to decode these clues.

There is also a second topic hidden below the surface in these lectures, i.e.,
how to make reliable calculations at low energy.  We will see that $V_{ud}$ is
known to 0.1\%, $V_{us}$ to 1\% and at least one mass ratio to 10\%.  For
the physics of hadrons these accuracies are remarkably good.  [For
example, $\alpha_s (M_z)$ is also only known to 10\%].  The key is the use
of symmetries as a dynamical tool.  In particular, we will be using chiral
perturbation theory.  While we do not have the time for a full pedagogical
presentation of this [1,2], we will see what it is and how it is used.

My approach here will reserve the heavy formalism as long as possible.  I
will treat quark masses crudely at first in order to get a basic feel for them
with a minimum of formalism.  Following that is the description of
$V_{ud}$.  Before proceeding on to describe the extraction of $V_{us}$, I will
spend some time
introducing chiral perturbation theory.  Finally I return to quark masses and
try to be as precise as possible.

\section{Quark Masses I}

Before turning to my real topic, we need to have a brief digression on
'constituent' vs. 'Lagrangian' or 'current' masses.  The Lagrangian of QCD

\begin{equation}
{\cal L}_{QCD} = - {1 \over 4} F^A_{\mu \nu} F^{A \mu \nu} +
\bar{\psi} (i D - m) \psi
\end{equation}

\noindent is a nonlinear field theory which contains small mass parameters
$m_u, m_d , m_s$.  Because these masses are small, the theory is
almost chirally symmetric, as well as almost classically scale invariant.
Masses also enter into the quark model of hadron structure, with

\begin{equation}
{\cal H}_{QM} = \sum_i {p^2_i \over 2M_i} + V(r_1 - r_2)
\end{equation}

\noindent Given that this is also supposed to represent the strong
interactions, it is remarkable how far this is from QCD.  The mass
parameters are large, \linebreak $M \sim m_p /3$, and there is no trace of the
symmetries of QCD.  The large mass of the quark model has very little
relation to the mass in the Lagrangian.  The former is commonly referred to as
a
'constituent' mass.  Our topic here concerns only the mass parameters in the
Lagrangian.  In many ways these are defined by the symmetry properties
and they are called 'current' (i.e., from divergences of Noether currents) or
'Lagrangian' masses.

Our first task is to learn to treat quark masses in the same way that we do
coupling constants.  Our mass parameters are not inertial masses of
hadrons, and because of confinement one cannot find any poles in quark
propagators.  How then can we come up with a way to actually measure
masses?  The procedure is the same as with coupling constants.
Observables depend on the masses, i.e.,

\begin{eqnarray}
M & = & M(m) \nonumber\\
& = & M_0 + am + bm^2  + \dots
\end{eqnarray}

\noindent We measure the quark mass m by its effect on observables.  But we
have a
problem; we cannot reliably calculate observables at low energy, and so it is
tough to learn how masses influence the observables.  It is here that
symmetry comes to the rescue.  There will be exact relations between
observables in the symmetry limit.  Quark masses break the symmetry and disturb
these relations.  That means that the deviations from the symmetry
predictions are measures of quark mass.  In the most basic of examples we
will see that the pion and kaon masses start off as

\begin{eqnarray}
m^2_{\pi} & = & 0 + (m_u + m_d) B_0 + \dots \nonumber\\
m^2_{K^+} & = & 0 + (m_u + m_s) B_0 + \dots
\end{eqnarray}

\noindent where $B_0$ is same constant.  This lets us measure the ratio

\begin{equation}
{m_u + m_d \over m_u + m_s} = {m^2_{\pi} \over m^2_{K^+}} + \dots
\end{equation}

\noindent This is the general plan for measuring quark masses [3].

Once we are treating masses as coupling constants, we are led to the issue
of renormalization.  If the Lagrangian is written in terms of bare parameters,
the interactions will induce mass shifts and we need to define renormalized
masses.  What then are the renormalization conditions and how are these
connected to observables?  I must admit that for the light quarks the answer
to this question has not been completely satisfactorily found at present.  In
perturbation
theory, of course, renormalization can be carried out.  However we do not have
a full connection between perturbation theory and
low energy measurements.  One key feature of perturbative renormalization
in QCD is that the mass shift of a fermion is proportional to the mass of that
fermion.  In general then we will find

\begin{equation}
m^{(R)}_i = m^{(bare)}_i \left[ Z_0 + Z_1 m_i + Z^{\prime}_1 \sum_{j \neq i}
m_i +
\ldots \right]
\end{equation}

\noindent To first order (in m) we always have

\begin{equation}
m^{(R)}_i \alpha \, m^{(bare)}_i
\end{equation}

\noindent so that ratios of the renormalized masses are equally ratios of the
bare parameters.  This nice feature can be preserved in mass independent
perturbative renormalization schemes.

In perturbative theory one can also choose to define running masses, $m_i
(q^2 )$.  In QCD, these get smaller as $q^2$ increases.  For light quarks
there is not much value for using these in the measurement of mass.  We
have our best information on ratios of masses, and in a mass independent
renormalization scheme, ratios are independent of the scale.  Another point
to be emphasized is that running masses for light quarks, despite getting
large at low $q^2$, do not make a good model for constituent masses.  This
is because all of the running masses vanish at all $q^2$ in the chiral limit
$(m^{(bare)}_i \rightarrow 0 \Rightarrow m_i (q^2) \rightarrow 0)$, in
contrast to constituent masses which approach a constant $(\approx 300
MeV)$ in this limit.

Non-perturbative effects can also induce mass shifts.  One possible new
form has been suggested by instanton calculations [4] with a mass shift

\begin{equation}
\delta m_u ~\alpha ~m_d m_s
\end{equation}

We will see later that this in fact is consistent with the symmetries of QCD.
It raises the question of what mass we are measuring in a given observable.
However let us save these issues for later and now turn to the simple lowest
order estimates of mass.

Consider first a world with massless $u, d, s$ quarks.  The quark helicity
(L, R) is not changed by QCD interactions in this limit, and is unchanged
under all Lorentz boosts.  There are then two separate worlds, with left
handed and right handed quarks being separately conserved.  This implies
an $SU(3)_L \times SU(3)_R$ symmetry.  Any mass will break this
symmetry because, at the very least, one can boost a massive left handed quark
to a
frame where it is right handed.  However, if $m$ is 'small' we are close to
the symmetry limit.  More precisely, in the massless limit, we have separate
global invariance under

\begin{eqnarray}
\psi_L \rightarrow \psi_L^\prime = L \psi_L \nonumber\\
\psi_R \rightarrow \psi_R^\prime = R \psi_R
\end{eqnarray}

\noindent with

\begin{equation}
\psi = \left( \begin{array}{c}
u \\ d \\ s
\end{array} \right)
\end{equation}

\noindent and L in $SU(3)_L$, R in $SU(3)_R$.  If there is a common
mass $m_u = m_d = m_s$, this chiral symmetry is explicitly broken to $SU(3)_V$,
and
separate masses for $u, d, s$ breaks even this latter symmetry.

However, while we see approximate $SU(3)_V$ multiplets in the spectrum
of hadrons, we do not see even approximate multiplets for $SU(3)_L \times
SU(3)_R$.  This is because the symmetry is hidden by the phenomena of
dynamical symmetry breaking.  This is characterized by a vacuum which is
not invariant under the symmetry, and the appearance of Goldstone bosons.
The $\pi, K, \eta$ are the Goldstone bosons, and would be massless if the
quarks were massless.  This fact can be used to yield the best known
measure of quark masses.  For it, we need to use only first order
perturbation theory, i.e., that the energy shift results from taking the matrix
element of the perturbing Hamiltonian between unperturbed wavefunctions.
The perturbation is

\begin{equation}
{\cal H}_m = m_u \bar{u}u + m_d \bar{d}d + m_d \bar{s}s
\end{equation}

\noindent and the results are

\begin{eqnarray}
< \pi \mid {\cal H}_m \mid \pi > & = & m^2_{\pi} = (m_u + m_d) B_0
\nonumber \\
< K^+ \mid {\cal H}_m \mid K^+ > & = & m^2_{K^+} = (m_u + m_s) B_0
\nonumber \\
< K^0 \mid {\cal H}_m \mid K^0 > & = & m^2_{K^0} = (m_d + m_s) B_0
\nonumber \\
< \eta \mid {\cal H}_m \mid \eta > & = & m^2_{\eta} = {1 \over 3} (4m_s +
m_u + m_d) B_0
\end{eqnarray}

\noindent where $B_0$ is a constant (the reduced matrix element).

Defining

\begin{equation}
\hat{m} = {m_u + m_d \over 2}
\end{equation}

\noindent we have

\begin{eqnarray}
{\hat{m} \over m_s} & = & {m^2_{\pi} \over 2m^2_K - m^2_{\pi}} \nonumber\\
{m_d - m_u \over m_s - \bar{m}} & = & {m^2_{K^0} - m^2_{K^+} \over
m^2_K - m^2_{\pi}}
\end{eqnarray}

\noindent valid to first order in the quark masses.  Actually the second of
these needs to be corrected for electromagnetic effects, which can also
influence the $K^0 - K^+$ mass difference.  Here we use Dashen's
theorem [5], i.e., that to lowest order in chiral SU(3) [that is, with no quark
mass  effects], the kaon and pion electromagnetic splitting are the same

\begin{equation}
(m^2_{K^+} - m^2_{K^0})_{EM} = m^2_{\pi^+} - m^2_{\pi^0}.
\end{equation}

\noindent Subtracting off this contribution leads to

\begin{equation}
{m_d - m_u \over m_s - \hat{m}} = {m^2_{K^0} - m^2_{K^+} -
m^2_{\pi^0} + m^2_{\pi^+} \over m^2_K - m^2_{\pi}} = 1/43
\end{equation}

\noindent or

\begin{equation}
{m_d - m_u \over m_d + m_u} = {m^2_{K^0} - m^2_{K^+} -
m^2_{\pi^0} + m^2_{\pi^+} \over m^2_{\pi}} = 0.28
\end{equation}

\noindent This is the estimate that most of the community is familiar with.
However, the full story on quark masses is considerably more involved (or
else my lectures would stop here).

Even at first order in the masses, there are other measures of quark mass
ratios.  Another interesting example is the decay $\eta \rightarrow 3 \pi$,
which is forbidden by isospin.  The electromagnetic effect vanishes at
lowest order in chiral SU(2) (Sutherland-Veltman theorem [6]) and all estimates
beyond this order indicate that electromagnetism has a negligible effect.  This
leaves the isospin breaking $m_d - m_u$ as the feature which induces
the decay.  Soft pion theorems can relate the amplitude to

\begin{equation}
\langle \pi^0 \mid {\cal H}_m \mid \eta \rangle = \sqrt{1 \over 3} (m_u - m_d)
B_0
\end{equation}

\noindent or the result can be read off from the effective Lagrangian
described later.   One finds

\begin{eqnarray}
{m_d - m_u \over m_d + m_u} & = & {3 \sqrt{3} \, F^2_{\pi} A_0
(\eta \rightarrow 3 \pi^+ \pi^- \pi^0) \over m^2_{\pi}} \nonumber \\
& = & 0.56
\end{eqnarray}

\noindent where $A_0$ is the amplitude in the center of the Dalitz plot and
the error bars are purely experimental.  This is considerably larger than the
previous result, and would imply $m_u/m_d = 1/3.5$.  However in
this case we do know some of the higher order effects (described later) are
sizeable, and will modify this result [7].  This result does indicate that
first
order measurements do not agree, and that we will need to confront the
analyses at second order.

A third measurement of quark masses at first order involves $\psi^\prime
\rightarrow J/\psi \pi^0$ and $\psi^\prime \rightarrow J/\psi \eta$.  The
former violates isospin and the second violates SU(3).  Again an
electromagnetism is estimated to play a very minor role, so that these decays
are driven by $m_d - m_u$ and $m_s - \hat{m}$ respectively.  The
analysis, using degenerate perturbative theory, yields the result

\begin{eqnarray}
{m_d - m_u \over m_s - \hat{m}} & = & \left[{16 \over 27} {\Gamma
\left(\psi^{\prime}
\rightarrow
J / \psi + \pi^0 \right) \over \Gamma \left( \psi^{\prime} \rightarrow  J /
\psi + \eta \right) }
{\rho^3_{\eta} \over P^3_{\pi}} \right]^{1/3} \nonumber \\
& = & 0.033 \pm 0.004
\end{eqnarray}

\noindent This calculation uses only vectorial SU(3), not chiral SU(3).  The
result lies almost exactly halfway between the answer given by meson
masses and by $\eta \rightarrow 3 \pi$ (which yields 0.023 and 0.046
respectively).  If we look at the spread around the central value, the first
order values have a standard SU(3) breaking spread of $1 \pm 30 \%$.

At this stage, one might ask about the absolute values of the masses.
However for the light quarks there is no measurement of the light quark
masses in the sense that I am using measurement.  The basic problem is that
the mass enters the theory multiplied by $\bar{\psi} \psi$, i.e., ${\cal H}_m
= m \bar{\psi}\psi$.  While their product is well defined, both m and
$\bar{\psi} \psi$ are separately renormalization scheme dependent, and the
measurements of the product do not measure m or $\bar{\psi} \psi$
separately.  A very rough determination is as follows [9].  Since $m_{u,d}
<< m_s$, we have at first order

\begin{eqnarray}
m_{\Lambda} - m_p & = & < \Lambda \mid {\cal H}_m \mid \Lambda > - <
P \mid {\cal H}_m \mid P > \nonumber \\
& \approx & < \Lambda \mid m_s \bar{s} s \mid \Lambda > - < P \mid m_s
\bar{s} s \mid P > \nonumber \\
& \equiv & m_s Z \nonumber \\
& \approx & 180 MeV
\end{eqnarray}

\noindent where

\begin{equation}
Z = \, < \Lambda \mid \bar{s} s \mid \Lambda > - < P \mid \bar{s} s \mid P >
\end{equation}

\noindent Because $< \Lambda \mid \bar{s} \gamma_0 s \mid \Lambda > =
1$, we might expect $Z \sim O(1)$.  [However, for the vacuum state $< 0
\mid \bar{s} \gamma_0 s \mid 0 > =0$ but $< 0 \mid \bar{s} s
\mid 0 >$ is quite large.]  Explicit quark model calculation [1] yields $Z =
0.5
\rightarrow 0.75$, which seem reasonable, but not extremely solid.  If we
use these we get

\begin{eqnarray}
m_s & \sim & 150 \rightarrow 300 MeV \nonumber \\
m_\alpha & \sim & 8 \rightarrow 16 MeV \nonumber \\
m_u & \sim & 3 \rightarrow 9 MeV
\end{eqnarray}

\noindent However, since these and other estimates of light quark masses
are based on models, not on measurements, we will not consider absolute
values further.

\section{The CKM Elements $V_{ud}, V_{us}$}

The weak mixing elements $V_{ud}$ and $V_{us}$ are best measured in
semileptonic decays, as nonleptonic transitions are not under theoretical
control.  The focus of theoretical analysis in the semileptonic decays is the
quest for precision in handling the strong interactions.  With $V_{ud}$, the
main issues are the electroweak radiative correction and small effects due to
isospin breaking.  For $V_{us}$, the primary concern is SU(3) breaking in
the current matrix elements.

The reference standard, to which the hadronic decays are compared, is
$\mu^- \rightarrow e^- \bar{\nu}_e \nu_{\mu}$.  With the
Hamiltonian

\begin{equation}
{\cal H}_w = {G_{\mu} \over \sqrt{2}} \bar{\nu}_{\mu} \gamma_{\mu}
(1 + \gamma_5 ) \mu \bar{e} \gamma^{\mu} (1 + \gamma_5 ) \nu_e
\end{equation}

\noindent and including the electroweak radiative correction, one has the
rate

\begin{eqnarray}
\Gamma (\mu \rightarrow e \nu \bar{\nu}) & = & {G^2_{\mu} m^5_{\mu} \over 192
\pi^3}
\left[1 -{\alpha \over 2 \pi} \left( \pi^2 - {25 \over 4}  \right) -
{8m^2_e \over m^2_{\mu}} +
{3 \over 5} {m^2_{\mu} \over  m^2_W} + \ldots \right] \nonumber \\
& = & {G^2_{\mu} m^5_{\mu} \over 192 \pi^3} \left[ 1 + (4203.85 -
187.12 + 1.05) \times 10^{-6} \right]
\end{eqnarray}

\noindent where the corrections, in the order written, are due to photonic
radiative effects, phase space, and the W propagator.  The value of
$G_{\mu}$ thus extracted is

\begin{equation}
G_{\mu} = 1.16637(2) \times 10^{-5} GeV^{-2}
\end{equation}

For $\Delta S = 0$ beta decays we have

\begin{equation}
{\cal H}_w = {G_{\beta} \over \sqrt{2}} V_{ud} \bar{u} \gamma^u (1 +
\gamma_5) d \bar{e} \gamma_{\mu} (1 + \gamma_5) \nu_e
\end{equation}

\noindent At tree level $G_{\mu} = G_{\beta}$, but at one loop this is no
longer true as there is an important difference in the radiative correction.
For the weak transition $1 + 3 \rightarrow 2 + 4$ some of the radiative
corrections are shown in Fig. 1.  Diagrams a, b are ultraviolet finite.  This
can
be understood by noting that the calculation is the same as the vertex
renormalization of a conserved current, which we know leads to no
renormalization at $q^2 = 0$.  Figures c, d are similar if we use the Fierz
transformation

\begin{equation}
\bar{\psi}_2 \gamma_{\mu} (1 + \gamma_5) \psi_1 \bar{\psi}_4
\gamma^{\mu} (1 + \gamma_5) \psi_3 = \bar{\psi}_4 \gamma_{\mu} (1 +
\gamma_5) \psi_1 \bar{\psi}_2 (1 + \gamma_5) \psi_1
\end{equation}

\noindent However diagrams e, f fall into a different class and are log
divergent if we use the Fermi interaction with no propagator.  The ultraviolet
portion is then proportional to $(Q_1 Q_3 + Q_2 Q_4)$, i.e.,

\begin{equation}
M^{(u.v.)}_{e,f} = -M^{(0)} {3\alpha \over 2\pi} (Q_1 Q_3 + Q_2 Q_4)
ln \Lambda / \mu_L
\end{equation}

\noindent where $\Lambda$ is a high energy cutoff and $\mu_l$ is a low
energy scale.  In muon decay $(1,2,3,4) = (\mu^-, \nu_{\mu}, \nu_e, e^-)$
so that

\begin{equation}
Q_{\mu} Q_{\nu_e} + Q_{\nu_\mu} Q_e = 0 .
\end{equation}

\noindent However in beta decay $(1,2,3,4) = (d, u, \nu_e, e^-)$ with

\begin{equation}
Q_d Q_{\nu_e} + Q_{u_\mu} Q_{e^-} = - {2 \over 3} .
\end{equation}

\noindent In a full treatment, including the $W$ propagator and $\gamma, Z$
loops one finds the cutoff $\Lambda = m_Z$, so that there is a universal
'model independent' correction [10] which can be absorbed in the definition of
$G_{\beta}$

\begin{equation}
G_{\beta} = G_{\mu} \left( 1 + {\alpha \over \pi} ln {M_Z \over \mu_L}
\right )
\end{equation}

\noindent To this also needs to be added smaller 'model dependent' low
energy effects and coulomb corrections.

For $\Delta S = 0$ decays, the key to mastering the strong interactions is
that the vector current is conserved (in the limit $m_u = m_d)$, so that the
matrix element is absolutely normalized.  In contrast it is not possible to
predict axial current matrix elements to high accuracy.  In neutron beta
decay, $n \rightarrow p e \nu $, where both vector and axial currents
contribute, one needs to measure the axial form factor $g_A$ in order to be
able to predict the rate and measure $V_{ud}$.  This works, but at present
the statistical accuracy is not the best.  Pion beta decay, $\pi^{\pm}
\rightarrow \pi^0 e^{\pm} \bar{\nu}$, only involves the vector current and
would be the ideal channel to study, but there are not yet enough events.  The
most sensitive process is $0^+ \rightarrow 0^+$ nuclear beta
decay between isospin
partners [11].  This also only involves the vector current, and has very high
statistics.

The superallowed $0^+ \rightarrow 0^+$ transitions have a single form
factor

\begin{equation}
< N_2 (I_z = 0) \mid V_{\mu} \mid N_1 (I_z = 1) > = a(q^2)(p_1 + p_2) .
\end{equation}

\noindent with $a(0) = \sqrt{2}$.  One calculates the half life $t_{1/2}$ times
a
kinematical phase space factor F, and adds hard and soft radiative
corrections, Coulomb corrections to the wavefunction and finite size effect

\begin{equation}
Ft_{1/2} = {2 \pi^3 ln 2 \over G^2_{\beta} m^5_e \mid V_{ud} \mid^2
a^2 (0)} [1 + \dots]
\end{equation}

\noindent Present efforts center on the nuclear wavefunction mismatch.
When one plots the Ft values for different nuclei vs. Z, the result should be
a constant value if all the nuclear effects have been taken into account
completely.  In practice there seems to be some indication for a slope to this
line [12], indicating that some effect linear in Z is not fully accounted for.
In the
recent analysis of Ref. 11 this has been corrected for phenomenologically be
extrapolating the Ft values to Z = 0, with the result

\begin{equation}
V_{ud} = 0.9751 \pm 0.0005
\end{equation}

\noindent One obtains a values for $V_{ud}$ about $2 \sigma$ lower if one
simply
averages the Ft measurements.  Neutron and pion beta decays are consistent
with Equation 35.

\section{Effective Lagrangian Description}

Before going on to the measurement of $V_{us}$, I need to describe the
uses of effective Lagrangian techniques in chiral perturbation theory.  In
these notes, I will be somewhat brief as Andy Cohen covers effective field
theory in these TASI lectures [2] and I have elsewhere [1] had the opportunity
to
present the subject in considerably greater depth.

The main idea is that if predictions follows from symmetry alone, then any
general Lagrangian with the right symmetry will yield the correct
predictions [13].  For physics of the light mesons, we seek then the most
general
Lagrangian with chiral SU(3) symmetry containing only the $\pi, K, \eta$
fields.  This can be accomplished with the $3 \times 3$ matrix representation

\begin{equation}
U = exp \left[ i {\vec{\lambda} \cdot \vec{\phi} \over F_{\pi}} \right] \; \; ,
\end{equation}

\noindent with transformation

\begin{equation}
U \rightarrow LUR^{\dagger}
\end{equation}

\noindent with L in $SU(3)_L$ and R in $SU(3)_3$.  The only Lagrangian
invariant
under chiral SU(3) with 2 derivatives (there are none with zero derivatives) is

\begin{equation}
{\cal L}_{SYM} = {F^2_{\pi} \over 4} Tr(\partial_{\mu} U
\partial^{\mu} U^{\dagger}) = {1 \over 2} \partial_{\mu} \phi^A \partial^{\mu}
\phi^A + \ldots
\end{equation}

\noindent For QCD we also need some explicit chiral symmetry breaking,
which at lowest order will be linear in the quark masses.  It preserves parity
and has the same chiral properties as $\bar{\psi} m \psi = \bar{\psi}_L m
\psi_R + \bar{\psi}_R m \psi_L$.  At lowest order the unique choice is

\begin{equation}
{\cal L}_{Breaking} = {F^2_{\pi} B_0 \over 2} Tr(m U + U^{\dagger}
m)
\end{equation}

\noindent where $B_0$ has been chosen to be the same constant as in
Section II.  The full lowest order Lagrangian

\begin{equation}
{\cal L} = {\cal L}_{sym} + {\cal L}_{Breaking}
\end{equation}

\noindent when applied at tree level reproduces all of the lowest order
predictions of chiral symmetry, such as the mass relations given previously.

What about effective Lagrangian with more derivatives or more powers of
the quark masses?  These may also have the correct chiral SU(3) properties.
The key to practical applications is the energy expansion.  Consider
two possible chirally symmetric Lagrangians

\begin{equation}
{\cal L}^1 = a Tr (\partial_{\mu} U \partial^{\mu} U^{\dagger} ) + b Tr
(\partial_{\mu} U \partial_{\nu} U^{\dagger} \partial^{\mu} U
\partial^{\nu} U^{\dagger})
\end{equation}

\noindent The Lagrangian has dimension $(mass)^4$, which implies that
$a$ has dimension $mass^2$ and $b/a \sim 1/mass^2$.  When matrix
elements are taken, derivatives turn into powers of momentum so that

\begin{equation}
M = a q^2 \left[1 + {b \over a} q^2 \right]
\end{equation}

\noindent If we define $b/a \equiv 1/\Lambda^2$, then for $q^2 \ll
\Lambda^2$ there is little effect of the higher derivative terms.  As $q^2$
increases, the four derivative term provides a correction to the lowest order
result.  In practice we most often find $\Lambda \sim m_{\rho}$, so that
lowest order chiral predictions are modified as momenta approach
$m_{\rho}$.

In constructing the effect of quark masses it is useful to consider an external
field of the form [14]

\begin{equation}
{\cal L}_{QCD} = \bar{\psi} i \rlap/{D} \psi - {1 \over 2B_0} \left(
\bar{\psi}_L \chi
\psi_R + \bar{\psi}_R \chi^{\dagger} \psi_L \right)
\end{equation}

\noindent such that QCD is obtained with $\chi = 2B_0 m$.  However, if we allow
a transformation rule

\begin{equation}
\chi \rightarrow L \chi R^{\dagger}
\end{equation}

\noindent the Lagrangian will be chirally invariant.  The effect of masses is
then found by writing chirally invariant Lagrangians containing $\chi$.  We
do this in Sec. 5.

Finally loop diagrams can, and must, be included.  Divergences appear, but
these just go into the renormalization of the parameters in the effective
Lagrangian.  Finite effects left over after renormalization account for the low
energy propagation of the pions and kaons.

The application of effective Lagrangians to the chiral interactions of $\pi, K,
\eta$ is called Chiral Perturbation Theory.  To next to leading order (i.e.,
$O(E^4)$) the instructions are:

\begin{enumerate}
\item Write the most general Lagrangians to $O(E^2)$ and $O(E^4)$; ${\cal L}_2$
contains two derivative or one power of the quark masses,  and ${\cal L}_4$
has either 4 derivatives, 2 derivatives and one mass, or two  powers of the
mass.
\item Calculate all one loop diagram involving ${\cal L}_2$
\item Renormalize the parameters in the Lagrangian, determining the
unknown parameters from experiment.
\item Find relations between different observables
\end{enumerate}

\noindent These relations are the predictions of chiral symmetry.

\section{$V_{us}$ and SU(3) Breaking}

One of the applications of chiral perturbation theory is in the determination
of $V_{us}$.
We will need to obtain the form factors in $\Delta S = 1$ processes
such as $K \rightarrow \pi e \nu$ and $\Lambda \rightarrow p e \nu$.
These are related by SU(3) to the $\Delta S = 0$ form factors which we
have already discussed $(\pi^+ \rightarrow \pi^0 e \nu, n \rightarrow p e
\nu)$.  However typical SU(3) breaking enters into other processes
at the 30\% level.  We
want to be more accurate than this.

A crucial ingredient here is the Ademollo Gatto theorem [15] which says that
the
vector form factors are modified from their SU(3) values only by terms
second order in the SU(3) breaking mass difference $m_s - \hat{m}$.  This
again points to the value of using vector form factors in the extraction of
$V_{us}$.  The two possible sources of data are hyperon decays and $K
\rightarrow \pi e \nu$.

Hyperon decays involve many modes and high statistics.  The axial form
factors cannot be predicted reliably from theory and must be measured.
SU(3) parameterizes these form factors in terms of two reduced matrix
elements, the neutron to proton axial coupling $g_A$ and a D/F ratio.
The vector form
factors are predicted via SU(3) plus the Ademollo Gatto theorem.  The
history of our ability to treat these decays has undergone fluctuations.
Before 1982, SU(3) fits worked well.  In 1982, the data improved enough
that SU(3) breaking at the 5\% level was observed and caused
troubles with fits based on SU(3)
symmetry, invalidating any fits using
SU(3) symmetry [16].  A few years later the quark model was used to provide
an SU(3) breaking pattern that was consistent with the data, allowing a
good fit and the extraction of $V_{us}$ [17].  Unfortunately by 1990, the data
was again better than theory, and the simple quark model pattern does not fit
without modification [18].  Unless theory can recover once again, hyperon
decays can not be analysed in any greater precision than this, because future
increased statistics will only tell us more details about SU(3) breaking.

Kaon semileptonic decays involves only two modes ($K^0$ and $K^+$
decay).  However the analysis is particularly strong since it can make use of
a body of work on chiral perturbative theory.  In addition these modes have
very high statistics.  For these reasons, kaon decay is the prime mode for
measuring $V_{us}$.

In order to be convinced that the theory of $K \rightarrow \pi e \nu$ is under
control, we have to turn to internal consistency checks.  The analysis is due
to Gasser and Leutwyler [19].  There are two form factors, $f_+$ and $f_-$

\begin{eqnarray}
< \pi^- \mid \bar{s} \gamma_{\mu} u \mid K^0 > & = & f^{K^0 \pi^-}_+ (k +
p)_{\mu} + f^{K^0 \pi^-}_- (k - p)_{\mu} \nonumber \\
< \pi^0 \mid \bar{s} \gamma_{\mu} u \mid K^+ > & = & {1 \over \sqrt{2}}
\left[ f^{K^+ \pi^0}_+ (k + p)_{\mu} + f^{K^+ \pi^0}_- (k - p)_{\mu}
\right].
\end{eqnarray}

\noindent If one includes the next-to-leading order Lagrangian, as well as
one loop diagrams, one obtains lengthy expressions for the form factors.
Among the highlights of the results are

\begin{enumerate}
\item The Ademollo Gatto theorem has a correction due to isospin breaking

\begin{eqnarray}
{f^{K^+ \pi^0}_+ (0) \over f^{K^0 \pi^-}_+ (0)} & = & 1 + {3 \over 4}
{m_d - m_{\mu} \over m_s - \hat{m}} + l_{K \pi} \nonumber \\
& = & 1.029 \pm 0.010 (Data)
\end{eqnarray}

\noindent where $l_{K \pi} = 0.004$ arise from loop diagrams.  This
value is consistent, because of the large uncertainty, with all of our previous
estimates of the quark mass ratio.

\item The form factors are related to the chiral constant $L_9$ determined in
the pion form factor,

\begin{eqnarray}
{f_- (0) \over f_+ (0)} & = & - \left[ 1 - {F_K \over F_{\pi}} + {2L_9 \over
F^2_{\pi}} \left(m^2_K - m^2_{\pi}\right) \right] \nonumber \\
& = & -0.13 \, theory \nonumber \\
& = & -0.20 \pm 0.08 \, data.
\end{eqnarray}

\item The slopes of the form factors are predicted in agreement with the data
(although the data presently have a few internal inconsistencies).

\end{enumerate}

\noindent Given that the theory appears to be under control Leutwyler and
Roos [20] have extracted

\begin{equation}
V_{us} = 0.2196 \pm 0.0023
\end{equation}

\noindent (a 10\% measurement).  This value is consistent with the results
of hyperon decay, and implies the check of the unitarity of the KM matrix

\begin{equation}
\mid V_{ud} \mid^2 + \mid V_{us} \mid^2 + \mid V_{ub} \mid^2 =
0.9990 \pm 0.0022
\end{equation}

\noindent with $\mid V_{ub} \mid^2 \leq 10^{-5}$.

\section{Quark Masses Beyond Leading Order}

Now we turn to the most difficult issue in these lectures; the analysis of
quark masses at second order.  There are several motivations for pursuing
such an analysis.  First of all, we have seen how the lowest order
predictions lead to some discrepancies.  In addition, there is the strong CP
problem [21], where the effect of CP violation by the $\theta$ term of QCD
would
vanish if $m_u \rightarrow 0$.  We then must question how well we
know that $m_u \neq 0$.  This solution to the strong CP problem is
not natural, in the technical sense, within the Standard Model, but perhaps
might be possible within an extension of our present theory.  Finally there
are several subtle issues which arise at second order in the mass, most
notably the reparameterization ambiguity described below.

The mass sector of the theory is described by

\begin{equation}
{\cal L} = \ldots + {F^2 \over 4} Tr \left( \chi^{\dagger} U + U^{\dagger}
\chi \right)
\end{equation}

\noindent at lowest order [recall Equation 43], and at higher order by

\begin{eqnarray}
{\cal L}_4 \ldots & + & L_6 \left[ Tr \left( \chi^{\dagger} U + U^{\dagger}
\chi \right) \right]^2 + L_7 \left[ TR \left( \chi^{\dagger} U - U^{\dagger}
\chi \right) \right]^2 \nonumber \\
& + & L_8 Tr \left( \chi U^{\dagger} \chi U^{\dagger} + \chi^{\dagger} U
\chi^{\dagger} U \right)
\end{eqnarray}

\noindent where $L_{6,7,8}$ are dimensionless unknown reduced matrix
elements in the basis of Gasser and Leutwyler [14].

The $\pi, K, \eta$ masses can be analysed to second order in the quark
masses [14].

\begin{eqnarray}
F^2_{\pi} m^2_{\pi} & = & 2 \hat{m} F^2 B_0 \left[1 + {32 L_6 B_0 \over F^2}
\left( m_u + m_d
+ m_s \right) + {32 L_8 B_0 \hat{m} \over F^2} \right. \nonumber \\
&   &\left. - 3 \mu_{\pi} - 2 \mu_K - {1 \over 3} \mu_{\eta} \right] \nonumber
\\
F^2_{K^+} m^2_{K^+} & = & \left( m_s + m_u \right) F^2 B_0 \left[ 1 + {32 L_6
B_0 \over F^2}
\left( m_u + m_d + m_s \right) \right. \nonumber \\
&   &\left. + {16 L_8 B_0 \over F^2} \left( m_u + m_s \right) - {3 \over 2}
\mu_{\pi} - 3 \mu_K - {5
\over 6} \mu_{\eta} \right] \nonumber \\
F^2_{K^0} m^2_{K^0} & = & \left( m_s + m_d \right) F^2 B_0 \left[ 1 + {32
L_6 B_0 \over F^2} \left( m_u  + m_d + m_s \right) \right. \nonumber \\
&   &\left. + {16 L_8 B_0 \over F^2} \left( m_s + m_d \right) - {3 \over 2}
\mu_{\pi} - 3 \mu_{K} -
{5 \over 6} \mu_{\eta} \right] \nonumber \\
F^2_{\eta} m^2_{\eta} & = & {4 \over 3} F^2_K m^2_K - {1 \over3} F^2_{\pi}
m^2_{\pi} -
{64 \over 3} \left( 2 L_7 + L_8 \right) B_0 \left( m_s - \hat{m} \right)^2
\nonumber \\
&   &+ \left( 2 \mu_{\pi} - {4 \over 3} \mu_K - {2 \over3} \mu_{\eta} \right)
\left( m^2_K -
m^2_{\pi} \right)
\end{eqnarray}

\noindent where

\begin{equation}
\eta_i = {m^2_i \over 32 \pi^2 F^2_\pi} ln m^2_i / \eta^2
\end{equation}

There are two main results of this analysis.  One is that the deviation of the
Gell Mann Okubo formula measures a useful combination of chiral
coefficients

\begin{eqnarray}
\delta_{GMO} & = & {4F^2_K m^2_K - 3F^2_{\eta} m^2_{\eta} - F^2_{\pi}
m^2_{\pi} \over 4 \left(F^2_K m^2_K - F^2_{\pi} m^2_{\pi} \right)}
\nonumber \\
& = & {16 \over F^2_{\pi}} \left(2L_7 + L_8 \right) \left( m^2_{\pi} -
m^2_K \right) - {3 \over 2} \mu_{\pi} + \mu_K + {1 \over 2} \mu_{\eta}
\nonumber \\
& = & -0.06 (Data)
\end{eqnarray}

\noindent which yields

\begin{equation}
2 L_7 + L_8 = 0.2 \times 10^{-3}
\end{equation}

\noindent at the chiral scale $\mu = m^2_{\eta}$.  The other prediction is
more important, producing a ratio of quark masses which are free from
unknown parameters

\begin{eqnarray}
{m_d - m_u \over m_s - \hat{m}} {2 \hat{m} \over m_s + \hat{m}} & = &{m^2_{\pi}
\over
m^2_K}
{\left( m^2_{K^0} - m^2_{K^+} \right)_{QM} \over m^2_K - m^2_{\pi}} \nonumber
\\
\left( m^2_{K^0} - m^2_{K^+} \right)_{QM} & \equiv & \left( m^2_{K^+} -
m^2_{K^+}
\right)_{expt}
- \left( m^2_{K^0} - m^2_{K^+} \right)_{EM}
\end{eqnarray}

\noindent The only flaw in this wonderful relation is that we do not know
$\left( m^2_{K^0} - m^2_{K^+} \right)_{EM}$ to the order that we are
working.  Recall that Dashen's theorem was only valid to zeroth order in the
quark mass.  The next order results have not been fully explored in chiral
perturbation
theory.

Gasser and Leutwyler have also analysed $\eta \rightarrow 3 \pi$ to second
order [7].  The result can be expressed in parameter free form as

\begin{equation}
{m_d - m_u \over m_s - \hat{m}} ~{2 \hat{m} \over m_s + \hat{m}} = {3
\sqrt{3} F^2_{\pi} Re A_{\eta \rightarrow 3 \pi} (0) \over \left[ 1 +
\Delta_{\eta 3 \pi} \right] \left( m^2_K - m^2_{\pi} \right)} ~{m^2_{\pi}
\over m^2_K} = 2.35 \times 10^{-3}
\end{equation}

\noindent where $\Delta_{\eta 3 \pi} = 0.5$.  Recall that this ratio was $1.7
\times 10^{-3}$ from meson masses and $3.5 \times 10^{-3}$ from $\eta
\rightarrow 3 \pi$, both at lowest order.  The effects of the $\em{O}(E^4)$
analysis has been to produce a compromise value for the ratio.

One of the advances of the past year is that it is now reasonable to expect
consistency between the analysis of the kaon mass difference and that of
$\eta \rightarrow 3 \pi$.  The agreement of Eq. 54 and Eq. 55 would require
$\left(
\Delta m^2_K \right)_{QM} = 7.0 MeV$.  However Dashen's theorem
implies $\left( \Delta m^2_K \right)_{EM} = 5.3 MeV$.  Are there
significant violations of Dashen's theorem?  Recent analyses suggest that
there are [22].  I am, of course, most partial to the work which I participated
in.  We used a series of powerful constraints on the $\gamma \pi \rightarrow
\gamma \pi$ and $\gamma K \rightarrow \gamma K$ amplitudes which
serve to predict the electromagnetic mass difference when the photons are
contracted into a propagator.  These constraints include 1) data on $\gamma
\gamma \rightarrow \pi \pi$, 2) low energy chiral constraints, 3) the
dispersion theory of $\gamma \gamma \rightarrow \pi \pi$, 4) soft pion
theorems and, 5) the generalized Weinberg sum rules.  These features are
compatible with a vector dominance model which yields

\begin{equation}
{\left( \Delta m^2_K \right)_{EM} \over \left( \Delta m^2_{\pi}
\right)_{EM}} = 1.8
\end{equation}

\noindent whereas Dashen's theorem says that the ratio should be unity.
The difference has a rather simple origin:  it is due to factors of $m^2_K$ in
propagators instead of $m^2_{\pi}$.  The larger electromagnetic contribution
brings the kaon mass difference
and $\eta \rightarrow 3 \pi$ into considerably better agreement (10 \%).

The most interesting feature of the study of quark masses beyond leading order
is the
reparameterization invariance, first made explicit by Kaplan and Manohar [23].
The crude
statement is that when using SU(3) symmetry one obtains the same physics using
either the masses
$(m_u , m_d , m_s )$ or the set

\begin{eqnarray}
m^{(\lambda)}_u & = & m_u + \bar{\lambda} m_d m_s \nonumber \\
m^{(\lambda)}_d & = & m_d + \bar{\lambda} m_u m_s \nonumber \\
m^{(\lambda)}_s & = & m_s + \bar{\lambda} m_u m_d
\end{eqnarray}

\noindent for an $\bar{\lambda}$!  The reason is that $m_i$ and
$m^{(\lambda)}_i$ both have the
same chiral SU(3) properties.  This can be seen using the Cayley-Hamilton
theorem for a $3 \times
3$ matrix A

\begin{equation}
det A = A^3 - A^2 Tr A - {A \over 2} \left[ Tr (A^2) - (Tr (A))^2 \right]
\end{equation}

\noindent If we apply this to the matrix $\chi$ [recall $\chi = 2 B_0 m$ for
pure QCD] defining

\begin{equation}
\chi^{(\lambda)} = \chi + \lambda [det \chi^{\dagger} ] \chi {1 \over
\chi^{\dagger} \chi}
\end{equation}

\noindent we have

\begin{eqnarray}
\left[ det \chi^{\dagger} \right] \chi \, {1 \over \chi^{\dagger} \chi} & = &
\left[ det U
\chi^{\dagger} \right] \chi \,{1 \over  \chi^{\dagger} \chi} \nonumber \\
& = & U \chi^{\dagger} U \chi^{\dagger} U - U \chi^{\dagger} U Tr \left( U
\chi^{\dagger} \right)  \nonumber \\
& - & {U \over 2} \left[ Tr \left( U \chi^{\dagger} U \chi^{\dagger} \right) -
\left( Tr U \chi^{\dagger} \right)^2 \right]
\end{eqnarray}

\noindent and

\begin{equation}
Tr (\chi^{\lambda} U^{\dagger}) = Tr (\chi U^{\dagger} ) - {\lambda \over 2}
\left[ Tr
(\chi^{\dagger} U \chi^{\dagger} U) - \left(Tr (\chi^{\dagger} U) \right)^2
\right]
\end{equation}

\noindent In an effective Lagrangian the use of $\chi^{(\lambda)}$ instead of
$\chi$ leads to a
Lagrangian of the same general form since

\begin{eqnarray}
Tr (\chi^{(\lambda)} U^{\dagger} + U \chi^{(\lambda)\dagger}) & = & Tr (\chi
U^{\dagger} + U
\chi^{\dagger}) \nonumber \\
& + & {\lambda \over 2} \left[ Tr (\chi U^{\dagger} + U \chi^{\dagger})
\right]^2 \nonumber \\
& + & {\lambda \over 2} \left[ Tr (\chi U^{\dagger} - U \chi^{\dagger})
\right]^2 \nonumber \\
& - & \lambda Tr (\chi U^{\dagger} \chi U^{\dagger} + \chi^{\dagger} U
\chi^{\dagger} U)
\end{eqnarray}

\noindent The last three terms lead to a modification of the chiral
coefficients which we called
$L_6, L_7, L_8$ previously.  However the total effective Lagrangian has the
same form.  Use of
$\chi^{(\lambda)}$ and one set of $L_6, L_7, L_8$ is equivalent to the use of
$\chi$ and a
different set of $L_6, L_7, L_8$.  This property of $\chi$ is the same as that
of the masses, when
we use $\chi = 2 B_0 m$, and

\begin{equation}
\chi^{(\lambda)} \equiv 2 B_0 m^{\lambda} = 2 B_0 \left[ m + (2 B_0 \lambda)
m_u m_d m_s {1
\over m} \right]
\end{equation}

\noindent and identify $\bar{\lambda} = 2 B_0 \lambda$.  The precise statement
of the
reparameterization ambiguity is then that, using either SU(3) or chiral SU(3)
any physics described
by $(m_u, m_d, m_s)$ and $(L_6, L_7, L_8)$ can be equally well described by

\begin{eqnarray}
m^{(\lambda)}_u = m_u + \bar{\lambda} m_d m_s &  L^{(\lambda)}_6 =
L_6 - \tilde{\lambda} \nonumber \\
m^{(\lambda)}_d = m_d + \bar{\lambda} m_u m_s &  L^{(\lambda)}_7 =
L_7 - \tilde{\lambda} \nonumber \\
m^{(\lambda)}_s = m_s + \bar{\lambda} m_u m_d &  L^{(\lambda)}_8 =
L_8 - 2 \tilde{\lambda} \end{eqnarray}

\noindent with $\bar{\lambda} = 2 B_0 \lambda ; \tilde{\lambda} = F^2_{\pi}
\lambda / 16$, for
any  reasonable $\lambda$.

Let us see examples of how this works.  For the ratio of quark masses measured
above, we have

\begin{eqnarray}
{m^{(\lambda)}_d - m^{(\lambda)}_u \over m^{(\lambda)}_s -
\hat{m}^{(\lambda)}} \,
{2 \hat{m}^{(\lambda)} \over {m^{(\lambda)}_s - \hat{m}^{(\lambda)}}}
& = & {(m_d - m_u)
(1 - \bar{\lambda} m_s) \over (m_s - \hat{m})(1 - \bar{\lambda} \hat{m})} \, {2
\hat{m} (1 +
\bar{\lambda} m_s) \over (m_s + \hat{m})(1 + \bar{\lambda} \hat{m})} \nonumber
\\
& = & {m_d - m_u \over m_s - \hat{m}}\, {2 \hat{m} \over m_s + \hat{m}} + {\cal
O} (m^2)
\end{eqnarray}

\noindent i.e., the ratio is invariant.  Similarly the combination

\begin{equation}
2 L^{(\lambda)}_7 + L^{(\lambda)}_8 = 2 (L_7 - \tilde{\lambda}) + (L_8 + 2
\tilde{\lambda}) =
2
L_7 + L_8
\end{equation}

\noindent is invariant.  Finally

\begin{eqnarray}
m^2_{\pi} & = & 2 B_0 \hat{m}^{(\lambda)} \left[ 1 + {32 B_0 \over F^2_{\pi}}
\left( \hat{m}
L^{(\lambda)}_8 + (2 \hat{m} + m_s) L^{(\lambda)}_6 \right) + \ldots \right]
\nonumber \\
& = & 2 B_0 \hat{m} (1 + \bar{\lambda} m_s) ]\left[ 1 - \bar{\lambda} m_s + {32
B_0 \over
F^2_{\pi}} \left( \hat{m} L_8 + (2 \hat{m} + m_s) L_6 \right) + \ldots \right]
\nonumber \\
& = & 2 B_0 \hat{m} \left[ 1 + {32 B_0 \over F^2_{\pi}} \left( \hat{m}
L_8 + (2 \hat{m} + m_s) L_6 \right) + \ldots \right] \nonumber \\
& & + {\cal O} (m^3)
\end{eqnarray}

\noindent is also unchanged in form under the reparameterization.

Physical quantities are invariant under the reparameterization transformation.
Quark mass ratios
(or the $L_i \, 's$) are not invariant and hence can not be uniquely measured
by any analysis
using  SU(3) or chiral SU(3).  This conclusion is general and extends to other
systems, such as
baryons  or heavy mesons, when analysed to second order (or beyond).  The best
that we can do is
to measure a one  parameter family of masses.

There is a weak restriction on the transformation in that we can't choose
$\lambda$ so large as to
destroy the energy expansion.  The typical sizes of the chiral coefficients are
of order a few times
$10^{-3}$.  We should not allow any $\tilde{\lambda}$ that makes $L_6, L_7,
L_8$ unnaturally
large.  In practice this does not happen for the mass range that we are most
interested in.

A conventional choice for masses and chiral parameters is

\begin{eqnarray}
{m_u \over m_s} = {1 \over 34} \; & ; & \; {m_d \over m_s} = {1 \over 19}
\nonumber \\
L_7 = -0.4 \times 10^{-3} \; & ; & \; L_8 = 1.1 \times 10^{-3}
\end{eqnarray}

\noindent A second set which is equally consistent is one with $m_u = 0$

\begin{eqnarray}
{m_u \over m_s} = 0 \; & ; & \; {m_d \over m_s} = {1 \over 26} \nonumber \\
L_7 = 0.2 \times 10^{-3} \; & ; & \; L_8 = -0.1 \times 10^{-3}
\end{eqnarray}

\noindent obtained by a reparameterization transformation.  A third compatible
set is

\begin{eqnarray}
{m_u \over m_s} = {1 \over 22} \; & ; & \; {m_d \over m_s} = {1 \over 16}
\nonumber \\
L_7 = -0.8 \times 10^{-3} \; & ; & \; L_8 = 1.9 \times 10^{-3}
\end{eqnarray}

\noindent In all cases $L_7$ and $L_8$ are natural in size.  (Nothing is known
about the
magnitude of $L_6$).  Note that since $m_u$ is the smallest mass, it changes
the most.  This is to
be expected since we have

\begin{equation}
\Delta m_u \sim m_d m_s \sim m_d {m^2_K \over \Lambda^2} \sim {1 \over 3} m_d
\sim m_u
\end{equation}

\noindent so that the change in $m_u$ is of the same order as $m_u$ itself.

The reparameterization transformation is an invariance of SU(3) effective
Lagrangians, not of the
fundamental QCD Lagrangian.  However, there may be physics in QCD which
generates effects
like this [24].  Let us consider the allowed forms of radiative corrections to
the masses in various
limits.

\begin{enumerate}
\item If $m_u = m_d = m_s = 0$, we have an exact $SU(3)_L \times SU(3)_R$
chiral symmetry.
There are no modifications to masses due to radiative corrections, as the
quarks are protected by
the chiral
symmetry from picking up a mass.
\item If $m_u = m_d = 0$ and $m_s \neq 0$, there is an exact chiral SU(2)
symmetry which
protects $m_u$ and $m_d$ from any quantum shifts. Likewise $m_u$ and $m_s$
would be
protected in an
$m_u = m_s = 0, m_d \neq 0$ world.
\item Now consider $m_u = 0$, but $m_d \neq 0$ and $m_s \neq 0$.  Now there is
no symmetry
protection at all, because chiral SU(2) is broken and axial U(1) is not a
quantum symmetry.
There can be radiative corrections to $m_u$.  However, since the corrections
must vanish as
$m_d \rightarrow 0$ or as $m_s \rightarrow 0$, it must have the form
\end{enumerate}

\begin{equation}
m_u = c m_d m_s
\end{equation}

\noindent There is in the literature an interesting example of just such a
renormalization, where
instantons lead to this form of radiative correction, with the overall
coefficient depending on the
cutoff in instanton sizes [4].  We don't need to take the details of this
calculation too seriously, but
we must acknowledge that this form of radiative correction can occur in QCD.
It is always
associated with the $U(1)_A$ anomaly.  By permutation symmetry, if $m_u \neq 0$
we would
have $\Delta m_d = c m_u m_s, \Delta m_s = c m_u m_d$.

These radiative corrections can produce different definitions of quark masses.
For example in a
mass independent renormalization scheme, one has

\begin{equation}
\left[ m^{(r)}_i \right]_1 = Z m_i
\end{equation}

\noindent with a common factor of Z.  In a second renormalization scheme one
might include the
low energy effects (such as the instantons) which induce the radiative
corrections of the preceding
paragraph.  The two schemes would be related by a finite renormalization

\begin{equation}
\left[ m^{(r)}_u \right]_2 = \left[ Z' m^{(r)}_u + \bar{\lambda} m^{(r)}_d
m^{(r)}_s \right]_1
\end{equation}

\noindent for some $\bar{\lambda}$.  For consistency, the various other
parameters in the theory
would also have to be related

\begin{equation}
\left[ L_7 \right]_2 = \left[L_7 - \tilde{\lambda} \right]_2
\end{equation}

\noindent such that observables are unchanged.  From this point of view, there
is the possibility of
a renormalization scheme ambiguity in QCD which mirrors the reparameterization
invariance.

An caveat to the above argument involves the $U(1)_A$ dependence.  In the
presence of a non-
zero vacuum angle $\theta$ in QCD the mass shift due to the instanton effect is
actually [4]

\begin{equation}
\Delta m_u = c m_d m_s e^{i \theta}
\end{equation}

\noindent The various masses of different renormalization schemes have
different $\theta$
dependence, and can in principle be differentiated by their behavior under
$U(1)_A$
transformations.  This can also be seen in the transformation of the $\chi$ and
$\chi^{(\lambda)}$
under $U(1)_A, L = e^{i \alpha}, R = e^{-i \alpha}$, in that

\begin{eqnarray}
\chi & \rightarrow & e^{2i \alpha} \chi \nonumber \\
\chi^{\lambda} & \rightarrow & e^{2i \alpha} \left[ \chi + \lambda e^{-6i
\alpha} \left[ det
\chi^+  \right] \chi {1 \over \chi^+ \chi} \right]
\end{eqnarray}

\noindent so that $m$ and $m^{(\lambda)}$ are not equivalent in their $U(1)_A$
properties.  Of
course, $U(1)_A$ is not a symmetry, but there are a set of anomalous Ward
identities [25] which
can in principle probe the $U(1)_A$ behavior.  In practice, none of the
measurements discussed
above involve $U(1)_A$.

There is an example which shows how the $U(1)_A$ properties can measure masses
independent
of the reparameterization [24] transformation.  Briefly summarized one adds the
$\theta F
\tilde{F}$ term to the QCD Lagrangian but with $\theta$ treated as an external
source so that
functional derivatives with respect to $\theta (x)$ yield matrix elements of $F
\tilde{F}$.  The
$U(1)_A$ properties determine how $\theta (x)$ enters the effective Lagrangian,
and these matrix
elements are calculated to $O(E^4)$.  The example shown was

\begin{eqnarray}
{< 0 \mid F \tilde{F} \mid \pi^0 > \over < 0 \mid F \tilde{F} \mid \eta >} = {3
\sqrt{3} \over 4}
\left[ {m_d - m_u \over m_s - \hat{m}} \right] {F_{\eta} \over F_{\pi}}
\nonumber \\
\left[ 1 - {32 B_0 \over F^2_{\pi}} (m_s - \hat{m}) (L_7 + L_8 ) + \ldots
\right]
\end{eqnarray}

\noindent This matrix element is not reparameterization invariant so that, if
it could be measured, it
could be used to disentangled the individual mass ratios.

What can be done in such a situation?  There is at present no completely
satisfactory solution.
However, some possible directions have been at least partially explored.  One
possibility is to
choose a definition of mass which is automatically reparameterization
invariant.  For example we
can simply define invariant masses $m^*_i$ by [24]

\begin{eqnarray}
F^2_{\pi} m^2_{\pi} & \equiv & F^2_0 B_0 \left[ m^{\ast}_u + m^{\ast}_d \right]
\nonumber \\
F^2_{K^+} m^2_{K^+} & \equiv & F^2_0 B_0 \left[ m^{\ast}_s + m^{\ast}_u \right]
+
\delta_{GMO} F^2_K \left( m^2_K - m^2_{\pi} \right) \nonumber \\
F^2_{K^0} m^2_{K^0} & \equiv & F^2_0 B_0 \left[ m^{\ast}_s + m^{\ast}_d \right]
+
\delta_{GMO} F^2_K \left( m^2_K - m^2_{\pi} \right) \nonumber \\
F^2_{\eta} m^2_{\eta} & \equiv & F^2_0 B_0 \left[ {4 \over 3} m^{\ast}_s + {2
\over 3}
\hat{m}^{\ast} \right]
\end{eqnarray}

\noindent This results in

\begin{eqnarray}
m^{\ast}_u & = & m_u \left[ 1 + {32 B_0 \over F^2_{\pi}}
\left( L_6 (m_u + m_d + m_s) + L_8 m_u \right) - 3 \mu_{\pi} - 2 \mu_K \right.
\nonumber \\
& & + \left.{1 \over 2} \left( {m_d - m_u \over m_s - \hat{m}} \right)
( \mu_{\eta} - \mu_{\pi}) \right] \nonumber \\
& & + \, 32 \, {L_7 B_0 \over F^2_{\pi}} (m_u - m_d) (m_u - m_s) \nonumber \\
\mu_i^2 & = & {m^2_i \over 32 \pi^2 \, F^2_{\pi}} \, ln \,{m^2_i / \mu^2}
\end{eqnarray}

\noindent with $m^{\ast}_d$ similar with $(m_u, m_d, m_s) \rightarrow (m_d,
m_u, m_s)$, and
$m^{\ast}_s$ likewise with $(m_u, m_d, m_s) \rightarrow (m_s, m_d, m_u)$, plus
some
rearrangement of the chiral logs [24].  Each of these invariant masses
$m^{\ast}_i$ is also
invariant
under changes in the scale $\mu$ which enters when using dimensional
regularization.  Many
ratios
of the $m^{\ast}_i$ are physical and can be evaluated

\begin{equation}
{m^{\ast}_d \over m^{\ast}_s} = {1 \over 22} \; ; \; {m^{\ast}_u \over
m^{\ast}_d} = 0.2
\end{equation}

\noindent and are fine measures of the breaking of chiral SU(3) and SU(2)
symmetry.  They do not
address the $U(1)_A$ properties and cannot answer the question of whether
strong CP violation
disappears due to
the $m_u = 0$ option.

A second possible direction is to try to use a model to calculate one of the
chiral coefficients.
Leutwyler has given a sum rule for $L_7$ and saturated it with an
$\eta^{\prime}$ pole [26].  This
is reasonable, but it is a model, and it is being applied in a sector where we
have no previous
experience to see if resonance saturation works in the presence of the
reparameterization
transformation.  As with many models, one often finds other contributions which
upset the original
conclusion -- as has been suggested for the $\pi^{\prime} (1300)$ intermediate
state in the sum
rule [27].

Ultimately the most promising way would be to find a way to measure observables
connected to
$U(1)_A$ anomalous Ward identities.  Wyler and I proposed to use $\psi^{\prime}
\rightarrow
J/\psi \pi^0$ and $\psi^{\prime} \rightarrow J/\psi \eta$ to do this [24, 28].
These unlikely
reactions
were chosen because an analysis by Voloshin and Zakharov [29] claimed that by
using a QCD
multipole [30] expansion, these decays were mediated by the local operator $F
\tilde{F}$, such
that a ratio of the decay rates can be converted into the ratio of Eq.(80).
This yielded a set of ratios
with $m_u \neq 0$.  However, the Voloshin Zakharov analysis has been criticized
by Luty and
Sundrum [31], and unless I am missing something it seems to me that the
criticism is justified.  I
have some hopes of getting around this problem in the future, but it is
otherwise difficult to
measure masses in $U(1)_A$ processes.

\section{Where do we stand?}

We have been using symmetries to measure masses and mixing angles.  The results

\begin{eqnarray}
V_{ud} & = & 0.9751 \pm 0.0005 \nonumber \\
V_{us} & = & 0.220 \pm 0.002 \nonumber \\
\mid V_{ud} \mid^2 + \mid V_{us} \mid^2 & = & 0.999 \pm 0.002
\end{eqnarray}

\noindent are gratifyingly precise.  For the expert the interest lies in the
error bars, which are
dominated by nuclear uncertainties in the case of $V_{ud}$, and SU(3) breaking
for $V_{us}$.

In the case of light quark masses, we have one firm ratio

\begin{equation}
{m_d - m_u \over m_s - \hat{m}} {2 \hat{m} \over m_s + \hat{m}} = 2.3 \times
10^{-3} ,
\end{equation}

\noindent accurate to about 10\%.  We cannot at present measure a second ratio
when we work
beyond leading order, due to the reparameterization transformation.  We are
left instead with a one
parameter family of mass ratios.  The up quark mass has the widest range,
presently including
$m_u = 0$.  Somewhat better known is $m_d / m_s \sim .05 (1 \pm 0.3)$.  More
precise
statements than this are model dependent.  In order to do better in the
measurement process, we
need to find a way to exploit axial $U(1)$ anomalous Ward identities.

\end{document}